# Generative AI Use in Entrepreneurship: An Integrative Review and an Empowerment–Entrapment Framework

**Authors:** Jackson G. Lu[1 *], Gerui Gloria Zhao[2], Anna Manyi Zheng[1]
**Affiliations:**
[1] MIT Sloan School of Management, Massachusetts Institute of Technology, Cambridge, MA, USA 02142
[2] School of Economics and Management, Tsinghua University, Beijing, China 100084

[*]Correspondence to Jackson G. Lu (lu18@mit.edu)

## Abstract

Despite the growing use of generative artificial intelligence (GenAI) in entrepreneurship, research on its impact remains fragmented. To address this limitation, we provide an integrative, entrepreneur-centered review of how GenAI influences entrepreneurs at each stage of the entrepreneurial process: (1) opportunity recognition and ideation, (2) opportunity evaluation and commitment, (3) resource assembly and mobilization, and (4) venture launch and growth. Based on our review, we propose the Empowerment–Entrapment Framework, which not only catalogs GenAI's benefits and costs throughout the entrepreneurial process, but also identifies potential trade-offs underlying GenAI's double-edged role. For example, GenAI may improve venture idea quality yet produce hallucinations and biases; boost entrepreneurial self-efficacy yet heighten overconfidence; increase functional breadth and self-sufficiency yet reduce prosociality; and enhance productivity yet diminish cognitive engagement, learning, and memory. We also identify features of GenAI that underlie these empowering and entrapping effects. Moreover, we explore boundary conditions that moderate GenAI's effects, including entrepreneurs' expertise, approaches to using GenAI, and personal traits and values. Beyond these theoretical contributions, our review and framework offer practical guidance for entrepreneurs seeking to use GenAI strategically while managing its risks.

**Acknowledgment:** We thank Celerina Chao, Jane Minyan Chen, Leslie Cheng, Christine Chiou, Blaire Han, Arthur Shiyun Hu, Xin Jin, Dana Kanze, Hannah Kwon, Ning Li, Sofia Li, Xiaoxuan Li, Yong Li, Hila Lifshitz, Yangduoduo Luo, Xufei Ma, Maya McPartland, Martin Obschonka, Kunyuan Qiao, Shuhua Sun, Yuqing Sun, Philipp von Sicherer, Juezihan Wang, Sophia Yutong Wang, Ruirui Rachel Yang, Mingrui Ye, Shaker Zahra, Lu Doris Zhang, Stephen Xu Zhang, and Eric Yanfei Zhao for their helpful feedback on earlier versions of the article.

## 1 | Introduction

Generative artificial intelligence (GenAI) is defined as "a category of AI that creates new content (such as text, images, audio, and video) by learning patterns from existing data" (Lu et al., 2025, p. 2360). Since the public release of tools such as ChatGPT, GenAI has become deeply integrated into work and life. In particular, GenAI is increasingly used in entrepreneurship, a context that requires rapid execution and continuous adaptation amid high uncertainty, pressure, and resource constraints (McMullen & Shepherd, 2006). Reflecting this trend, a survey of thousands of entrepreneurs found that 47% of those who founded businesses in 2024 used GenAI, more than double the share in 2023 (Gusto Insights Group, 2025). As entrepreneurs increasingly use GenAI, an important research question arises: How does GenAI use affect entrepreneurs at different stages of the entrepreneurial process?

The existing literature has not yet provided a clear answer due to two main limitations. The first is that the literature is fragmented with scattered studies examining isolated outcomes (e.g., idea quality, efficiency, and overconfidence) across disparate entrepreneurial activities. To address this limitation, we conduct an integrative review of GenAI's effects on individual entrepreneurs. Based on this entrepreneur-centered review, we develop a stage-based framework that maps GenAI's effects onto each stage of the entrepreneurial process (Table 1): (1) opportunity recognition and ideation, (2) opportunity evaluation and commitment, (3) resource assembly and mobilization, and (4) venture launch and growth.

The second limitation is that although existing studies have documented positive and negative effects of GenAI use, these findings have not been synthesized to explain when and why GenAI empowers or entraps entrepreneurs across different entrepreneurial activities. To address this limitation, we further develop the stage-based framework into the Empowerment–Entrapment Framework, which highlights GenAI's double-edged role at each stage of the entrepreneurial process (see Figure 1). We define *empowerment* as the expansion of an entrepreneur's capabilities to engage in entrepreneurial activities, and *entrapment* as the restriction or erosion of those capabilities. We also identify features of GenAI that underlie these empowering and entrapping effects, including its ability to complete tasks efficiently and conveniently, as well as its training data and design features. Moreover, we explore boundary conditions that moderate GenAI's effects, including entrepreneurs' expertise, approaches to using GenAI, and personal traits and values.

-----------------------------

Insert Figure 1 about here

-----------------------------

Beyond these theoretical contributions, our paper offers practical guidance for entrepreneurs seeking to use GenAI effectively while managing its risks. In particular, we highlight the importance of recognizing and navigating the trade-offs between GenAI's empowering and entrapping effects, calibrating its use strategically, protecting relational assets throughout the entrepreneurial process, and determining what tasks to delegate to GenAI.

## 2 | Literature Search

To ensure the comprehensiveness of our review, we followed a structured and multistep literature search process. Since ChatGPT was publicly released on November 30, 2022, we limited our review to studies written in English from 2023 to July 2026, a period characterized by the rapid rise of GenAI tools and a corresponding surge in research on their applications in entrepreneurship and other workplace settings.

First, we searched major academic databases, including the Web of Science Core Collection, Scopus, Business Source Complete, and APA PsycInfo. We used combinations of entrepreneurship-related keywords (e.g., “entrepreneur*,” “founder*,” “startup,” “start-up,” “new business,” and “new venture”) and GenAI-related keywords (e.g., “generative artificial intelligence,” “generative AI,” “large language model,” “LLM,” “chatbot,” “GPT,” “Claude,” “Copilot,” “Gemini,” “Llama,” “DeepSeek,” “Doubao,” and “ERNIE”).

Second, because research on GenAI and entrepreneurship is still emerging, we also conducted supplementary searches in Google Scholar to identify recent preprints, including those not yet indexed in the academic databases searched above. Using combinations of the entrepreneurship- and GenAI-related terms listed above, we reviewed the first 50 results returned for each keyword combination (Qin, Zhou, et al., 2025).

Third, we conducted backward citation searches by reviewing the reference lists of the included articles and forward citation searches by identifying subsequent articles that cited them. This process allowed us to identify additional relevant studies, including research conducted in adjacent contexts but relevant to entrepreneurial activities.

After removing duplicates, 836 unique records remained for title and abstract screening. We applied four exclusion criteria. First, we excluded articles that did not address our focal question of how GenAI influences entrepreneurs at the individual level. Specifically, we excluded articles focused solely on *industry-level* or *macroeconomic* outcomes (e.g., market concentration, industry productivity, or regional innovation effects). Second, we excluded articles examining non-GenAI (e.g., conventional machine-learning systems, medical diagnosis applications, or AlphaGo) rather than GenAI (e.g., large language models [LLMs]). Third, we excluded purely technical papers focused on algorithmic architectures, model training methods, or computational performance without implications for entrepreneurial activities. Fourth, we excluded editorials, book reviews, opinion pieces, and sources lacking sufficient methodological transparency or theoretical relevance.

Following title and abstract screening and full-text assessment based on these criteria,

we retained 112 articles for the final review. These included studies that directly examined GenAI in entrepreneurial settings, as well as studies that examined GenAI's effects in adjacent contexts relevant to entrepreneurial activities, such as creativity, learning, and productivity (Table S1). Although studies from adjacent contexts did not always involve entrepreneurs, their findings provide theoretically relevant insights into how GenAI may affect activities that are central to the entrepreneurial process. Throughout the review, we distinguish between these two complementary streams of evidence, synthesize their findings, and use calibrated language when extending findings from adjacent contexts to entrepreneurship. We also call for future research to examine whether, and under what conditions, effects documented in adjacent contexts generalize to entrepreneurial settings.

## 3 | The Empowerment–Entrapment Framework: GenAI's Benefits and Costs Across Entrepreneurial Stages

Our Empowerment–Entrapment Framework has four theoretical features. First, it is a *process* model that conceptualizes entrepreneurship as unfolding across multiple stages. Although scholars differ in the labels and granularity used to describe this process, *process* perspectives generally emphasize interconnected activities through which individuals identify potential opportunities, evaluate and commit to selected opportunities, mobilize resources, and bring ventures to market (Baron, 2007b; Baron & Shane, 2007; Bhave, 1994; Davidsson & Gruenhagen, 2021; Gartner, 1985; Shepherd et al., 2021).

We draw on this process literature to organize our review around four stages of the entrepreneurial process: (1) opportunity recognition and ideation, (2) opportunity evaluation and commitment, (3) resource assembly and mobilization, and (4) venture launch and growth. Table 1 summarizes the key entrepreneurial tasks associated with each stage.

-----------------------------

Insert Table 1 about here

-----------------------------

Importantly, while our framework presents the entrepreneurial process linearly (i.e., as a sequence of stages)—an approach commonly used in the literature (Baron, 2007b; Baron & Shane, 2007)—we recognize that the entrepreneurial process may unfold in a nonlinear, iterative manner, as emphasized in the effectuation literature and related perspectives (e.g., Bhave, 1994; Leatherbee & Katila, 2020; Sarasvathy, 2001). For structural clarity, however, our framework adopts a linear representation of the entrepreneurial process to delineate the benefits and costs of GenAI use, while recognizing that many of these benefits and costs are likely to apply across both linear and nonlinear processes.

Second, the Empowerment–Entrapment Framework delineates the benefits and costs of GenAI use at each stage of the entrepreneurial process (Sections 3.1 to 3.4). Beyond merely cataloging these benefits and costs, the framework illuminates their potential trade-offs to highlight GenAI's double-edged role. For example, GenAI may boost entrepreneurial self-efficacy while heightening overconfidence; enhance productivity while diminishing cognitive engagement, learning, and memory; and expand functional breadth and increase self-sufficiency while reducing prosociality.

Notably, while some effects are most directly tied to the focal entrepreneurial task of a particular stage (e.g., improved idea quality, efficient feasibility testing, algorithmic biases in hiring, and premature venture launch), other effects recur across multiple stages of the entrepreneurial process (e.g., hallucinations, training-data biases, increased entrepreneurial self-efficacy, and diminished cognitive engagement, learning, and memory). For structural clarity, we discuss each cross-stage effect primarily at the stage where it is especially relevant, but the effect should not be interpreted as confined to that stage.

Third, the framework identifies core features of GenAI that provide common bases for these benefits and costs (Section 3.5).

Fourth, the framework discusses boundary conditions that moderate GenAI's effects (Section 3.6).

## 3.1 | Stage 1: Opportunity Recognition and Ideation

The entrepreneurial process begins with opportunity recognition, during which entrepreneurs identify problems, unmet needs, or inefficiencies that may create opportunities for new ventures (Ardichvili et al., 2003; Baron, 2006). Through ideation, entrepreneurs generate ideas for products or services to address the identified opportunities (Dimov, 2007).

### 3.1.1 | Improved idea quality vs. Hallucinations and training-data biases.

***3.1.1.1 | Improved idea quality.*** A growing number of studies suggest that GenAI can enhance the quality of venture ideas during ideation. For example, an experiment found that business ideas generated through human–GenAI collaboration outperformed those developed by humans, demonstrating superior overall quality, as rated by external human evaluators based on criteria such as strategic viability, financial value, and environmental value (Boussioux et al., 2024). Similarly, a field experiment on product innovation found that *individuals with GenAI assistance* generated solutions that both domain experts and GenAI rated as comparable in overall quality to those generated by *two-person teams without GenAI assistance* (Dell'Acqua, Ayoubi, et al., 2026). Moreover, without GenAI, research and development (R&D) professionals tended to propose technically oriented solutions, whereas commercial professionals tended to propose commercially oriented solutions. With GenAI, however, professionals from both backgrounds proposed more integrative solutions that effectively balanced technical and commercial considerations (Dell'Acqua, Ayoubi, et al., 2026). Applied to the entrepreneurial context, these findings suggest that GenAI may be especially beneficial for solo or novice founders, who often lack the cross-domain expertise necessary to generate high-quality venture ideas.

***3.1.1.2 | Hallucinations.*** Despite these benefits, GenAI is also prone to hallucinations, defined as outputs that appear plausible but contain fabricated or unsupported information (Ciubotaru, 2025; Farquhar et al., 2024; Saeri et al., 2026). For example, GenAI has fabricated legal precedents in court filings and generated false factual claims in news articles (Bohannon, 2023; Opdahl et al., 2023).

In the entrepreneurial context, GenAI hallucinations may be especially consequential during the ideation stage, when entrepreneurs may use GenAI to conduct preliminary market research, identify unmet needs, and form early beliefs about the viability of a potential opportunity. If left unchecked, hallucinated content may become embedded in entrepreneurs' initial ideas and propagate through later stages of the entrepreneurial process. This propagation may amplify downstream risks by reinforcing an illusion of objectivity during opportunity evaluation (see Section 3.2.1.2) and an illusion of readiness during venture launch (see Section 3.4.1.2).

***3.1.1.3 | Training-data biases.*** Another potential cost of using GenAI is the systematic biases reflected in its outputs. Since the data used to train GenAI can disproportionately represent certain regions, languages, and cultures, GenAI outputs may systematically reproduce these imbalances, often privileging Western perspectives (Gallegos et al., 2024; Tao et al., 2024). For example, in a study comparing GenAI estimates of moral values across 48 countries with survey responses from 90,802 human participants, Zewail et al. (2026) found that GPT models overestimated the overall moral concerns of some Western countries (e.g., Australia, Canada, and the U.S.) while underestimating those of some non-Western countries (e.g., Indonesia, Morocco, and Nigeria). Relatedly, Manvi et al. (2024) examined five GenAI models (GPT-4 Turbo, GPT-3.5 Turbo, Gemini Pro, Mixtral 8×7B, and Llama 2 70B) and found that they all exhibited biases against regions with lower socioeconomic conditions (e.g., most of Africa) on sensitive and subjective topics such as

attractiveness, morality, and intelligence.

These distortions are not limited to cultural or geographic representation; they also extend to demographic representation. Santurkar et al. (2023) found a substantial misalignment between the opinions reflected in GenAI outputs and those held by diverse U.S. demographic groups. Specifically, GenAI models poorly represented the views of some groups (e.g., adults aged 65 or older or widowed individuals), and this misalignment persisted even when GenAI was explicitly prompted to adopt those groups' perspectives (Santurkar et al., 2023). One reason is that GenAI's training data are often drawn from internet sources that overrepresent groups with greater digital access and higher participation rates and privilege viewpoints that are more visible on dominant platforms. For example, the training data are sometimes drawn from sources such as Reddit, whose user base skews young and male, and Wikipedia, whose editor base is also predominantly male (Bender et al., 2021). Moreover, data collection and filtering practices may further suppress the voices of marginalized groups (Bender et al., 2021).

In sum, GenAI may generate ideas that appear broadly useful while in fact reflecting the preferences, assumptions, and experiences of certain groups. This poses a particular risk during entrepreneurial ideation: When using GenAI, entrepreneurs may gravitate toward opportunities that resonate with certain audiences while overlooking unmet market needs among marginalized or less visible populations.

**3.1.2 | Increased creativity vs. Decreased distinctiveness of venture ideas.**

***3.1.2.1 | Increased creativity.*** Creativity lies at the core of the ideation stage of entrepreneurship (Baron & Shane, 2007). A growing body of research demonstrates that using GenAI enables individuals to produce more creative ideas than they do without GenAI (Lee & Chung, 2024; Mei et al., 2025; Sun et al., 2025; Zhu & Zou, 2026). For example, across five experiments, Lee and Chung (2024) found that individuals using GenAI generated

more creative ideas than those using either no technology or a conventional web search. Similarly, in a field experiment at a technology consulting firm, Sun et al. (2025) found that employees receiving GenAI assistance exhibited greater creativity than those without GenAI assistance, as assessed by both supervisors and external evaluators.

One mechanism underlying GenAI's empowering effect on creativity is its capacity to expand individuals' *cognitive resources*, defined as resources essential for addressing the cognitive demands of tasks (Sun et al., 2025). Creativity requires recombining diverse knowledge components in novel and useful ways (Hargadon, 2002; Lu, Akinola, et al., 2017; Lu, Hafenbrack, et al., 2017), but acquiring and processing such knowledge is cognitively demanding (Jones, 2009). GenAI can support this process by facilitating access to information about emerging trends, market gaps, and innovative business strategies, thereby expanding the knowledge available for recombination (Duong & Nguyen, 2024; J. Kim, 2025; Sun et al., 2025). GenAI can also free cognitive resources for creative work by handling routine tasks such as summarization and preliminary research (Lu, Sun, et al., 2026). Supporting this cognitive-resource mechanism, Sun et al.'s (2025) field experiment found that cognitive resources mediated the effect of GenAI use on creativity.

Notably, the effects of GenAI on creativity are not uniform across individuals. One moderator is individuals' metacognition, defined as "actively monitoring and regulating one's thinking to complete tasks and achieve goals" (Sun et al., 2025, p. 1564). For example, Sun et al.'s (2025) field experiment found that while GenAI use enhanced creativity on average, this effect was more pronounced for individuals with higher metacognition. Entrepreneurs with higher metacognition may be more likely to consider how best to interact with GenAI and monitor whether AI-generated ideas meaningfully expand the opportunity space. By contrast, entrepreneurs with lower metacognition may be more likely to rely on GenAI's initial suggestions and engage in less iterative exploration, thus limiting the extent to which GenAI enhances their creativity.

In addition, studies suggest that GenAI tends to yield greater creative gains for individuals with lower baseline creativity (Doshi & Hauser, 2024) or lower domain expertise (M. Huang et al., 2024). These findings suggest that GenAI use may be particularly beneficial for entrepreneurs who lack deep industry knowledge or prior entrepreneurial experience. Furthermore, entrepreneurs operating across linguistic or cultural boundaries, such as *non-native speakers* engaging with global markets, may benefit more from GenAI's support in idea development and articulation, because GenAI can reduce language barriers and improve expressive clarity (Mei et al., 2025).

***3.1.2.2 | Reduced distinctiveness of venture ideas.*** Although GenAI may enhance entrepreneurs' creativity relative to what they could achieve without GenAI, the ideas they generate may be less distinctive than those generated by other entrepreneurs. In an experiment, Doshi and Hauser (2024) asked participants to write stories and found that while GenAI-assisted stories were rated as more creative, they were also more similar to one another than stories written without GenAI. Similarly, E. Zhou and Lee (2024) tracked over four million digital artworks from more than 50,000 users and found that after artists adopted text-to-image GenAI tools (e.g., Midjourney, Stable Diffusion, DALL·E), their artworks became increasingly similar on average in their focal objects, themes, and visual styles over time. Likewise, Anderson et al. (2024) found that, compared with a non-AI creativity support tool (the Oblique Strategies deck), GenAI-assisted ideation increased the number of ideas generated by individuals but produced more homogeneous idea sets across users at the group level. In entrepreneurial settings, Cheng and Gao (2026) analyzed 291,314 product launches on Product Hunt and found that startup ideas became more similar after ChatGPT's release, with the share of distinctive launches falling from 25% to 16%, particularly in categories where GenAI could more easily assist ideation. Hunt and Kurdoglu (2025) argued that GenAI

use can reinforce convergence in entrepreneurial ideation by filtering out anomalous ideas and steering entrepreneurs toward similar opportunity spaces.

One reason for this convergence is that many GenAI models are trained on overlapping large-scale corpora (e.g., Common Crawl) and generate outputs by predicting successive tokens based on patterns learned from training data (Brown et al., 2020; Gao et al., 2020; Wang et al., 2026). As a result, different entrepreneurs providing comparable inputs may receive similar outputs (Doshi & Hauser, 2024; Meincke et al., 2025; Y. Zhou et al., 2026). Such convergence may make it more difficult for an entrepreneur to differentiate their venture and establish a distinctive market position.

Importantly, this convergence does not contradict the creative benefits of GenAI discussed in Section 3.1.2.1. GenAI may broaden the range of ideas available to each entrepreneur relative to their own baseline while simultaneously making ideas across entrepreneurs more similar—that is, increasing *within-person* creativity while reducing *between-person* distinctiveness.

**3.1.3 | Stage 1 summary.** Supporting our Empowerment–Entrapment Framework (Figure 1), GenAI may function as a double-edged sword during the opportunity recognition and ideation stage. While GenAI may improve the quality of venture ideas, its hallucinations and training-data biases may also harm venture ideas. In addition, while GenAI may increase entrepreneurs' creativity, reliance on similar GenAI models may make their ideas less distinctive relative to those developed by competitors. These early-stage effects may be consequential because they can shape entrepreneurs' beliefs about potential opportunities, the directions they explore, and the ideas they carry into later stages of the entrepreneurial process.

## 3.2 | Stage 2: Opportunity Evaluation and Commitment

The opportunity evaluation and commitment stage involves systematically assessing the viability of opportunities and venture ideas, conducting feasibility analyses, and gathering feedback to better understand potential risks and rewards (Ardichvili et al., 2003; McMullen & Shepherd, 2006; Wood & Williams, 2014). Based on this evaluation, entrepreneurs decide whether to commit resources to pursuing the opportunity, defer action while gathering additional information, or abandon it altogether (Baron & Shane, 2007; McMullen & Shepherd, 2006).

#### 3.2.1 | Structured and informative evaluation vs. Illusion of objectivity.

***3.2.1.1 | Structured and informative evaluation.*** GenAI may streamline opportunity evaluation by providing entrepreneurs with decision-making frameworks (e.g., logic trees and comparison tables) and systematically assessing venture opportunities and ideas. Prior research suggests that GenAI can support strategic decision-making by gathering relevant information and offering structured assessments of various alternatives (Albashrawi, 2025; Alkayyal et al., 2026; Handler et al., 2024). This structured approach may help entrepreneurs evaluate opportunities more systematically and make more informed decisions. For example, an entrepreneur could use GenAI to create a comparison table of potential solutions to sleep disorders among college students (e.g., a wearable tracker vs. a coaching app), assessing each option on dimensions such as competition, resource requirements, and time to profitability, thereby facilitating a structured evaluation of trade-offs. Beyond providing structure, emerging evidence suggests that AI-generated evaluations may align with investor judgments. Csaszar et al. (2024) found that GenAI's evaluations of startup business plans were positively correlated with those of *experienced* venture capital and angel investors across dimensions such as market opportunity, competitive advantage, investment potential, and viability. Similarly, Kleinert and Urbig (2026) found that AI-generated investment evaluations successfully differentiated between ventures that subsequently succeeded in raising funds and those that did not.

***3.2.1.2 | Illusion of objectivity.*** However, GenAI may also create an illusion of objectivity. Because GenAI often presents information in structured formats, includes data and references, and adopts a confident tone, its outputs may appear analytically rigorous and evidence-based. Yet, as noted in Sections 3.1.1.2 and 3.1.1.3, these outputs may contain hallucinations or biases. Because such problems can be difficult to detect, entrepreneurs may infer credibility from the apparent rigor, mistake cogency for accuracy, and overweight AI-generated assessments (D'Angelo et al., 2026). Relatedly, users' confidence in AI-generated answers may be poorly calibrated to the answers' actual accuracy (Steyvers et al., 2025). For example, in a study of problem-solving, Krupp et al. (2024) found that participants frequently misjudged the correctness of AI-generated solutions, with more than half of incorrect responses being accepted as correct. Similarly, D'Angelo et al. (2026) found that entrepreneurs could be impressed by GenAI's detailed, coherent, and seemingly well-reasoned responses, potentially treating them as more objective and accurate than warranted. In the context of opportunity evaluation, this illusion of objectivity may lead entrepreneurs to underinvest in independent verification (e.g., external data, customer feedback, and market validation) and weaken critical scrutiny of whether an opportunity is truly viable.

**3.2.2 | Efficient feasibility testing vs. Information overload.**

***3.2.2.1 | Efficient feasibility testing.*** GenAI may help entrepreneurs conduct feasibility testing and experimentation efficiently during opportunity evaluation. In social science research, GenAI may support experimentation by helping develop study designs (Eymann et al., 2025; Park et al., 2024), forecasting experimental results (Ashokkumar et al., 2026; Cui et al., 2025), and enabling hypothesis testing at a fraction of the cost of studies involving human participants (Qin, Huang, et al., 2025). Moreover, Bilgram and Laarmann (2023) showed that GenAI can help users (including those without technical expertise) translate product ideas into early digital prototypes, thereby facilitating early concept

validation and feasibility testing. Similarly, a study across nine startups found that GenAI can lower the cost of experimentation and compress the idea-to-test cycle, enabling multiple tests to be conducted in parallel (Santoro et al., 2026). These findings suggest that GenAI can help entrepreneurs translate venture ideas into testable hypotheses and design small-scale tests, thereby enabling faster and less costly feasibility testing before they commit substantial resources (Ardichvili et al., 2003).

***3.2.2.2 | Information overload.*** However, GenAI may also generate an overwhelming amount of information for entrepreneurs to interpret. Given the limits of human cognitive capacity (Miller, 1956), processing numerous alternatives and detailed outputs can make it cognitively taxing for entrepreneurs to compare trade-offs and identify the most promising opportunity. Under such conditions, entrepreneurs may be more likely to rely on intuitive, heuristic-based reasoning rather than deliberative, analytical reasoning (Kahneman, 2011). Consistent with this reasoning, an experiment on strategic decision-making found that GenAI use increased information overload during startup evaluation (Kanis et al., 2026). D'Angelo et al. (2026) similarly identified information overload as a key challenge arising from GenAI use, as entrepreneurs struggled to process the large volume of AI-generated information and refine their ideas. During opportunity evaluation, GenAI may present entrepreneurs with too many alternatives, making it difficult for them to narrow their choices and commit to a focal opportunity.

**3.2.3 | Entrepreneurial self-efficacy vs. Entrepreneurial overconfidence.**

***3.2.3.1 | Entrepreneurial self-efficacy.*** Entrepreneurs often face uncertainty when navigating ambiguous markets and other unfamiliar challenges (Ramoglou et al., 2025). In this context, GenAI may enhance entrepreneurial self-efficacy, defined as "a person's belief in their ability to successfully launch an entrepreneurial venture" (McGee et al., 2009, p. 965). By providing timely information, structured guidance, and feedback during opportunity

evaluation, GenAI may help entrepreneurs better understand the decision context and relevant considerations and may make unfamiliar entrepreneurial tasks appear more manageable (Y. Li et al., 2025; Susskind et al., 2024). Consistent with this possibility, entrepreneurship research shows that GenAI adoption is positively associated with entrepreneurial self-efficacy, which in turn predicts stronger entrepreneurial intentions (Bui & Duong, 2024; Duong & Nguyen, 2024; Xie & Wang, 2025) and greater engagement in entrepreneurial activities (Duong & Nguyen, 2024).

***3.2.3.2 | Entrepreneurial overconfidence.*** However, GenAI may also induce overconfidence among entrepreneurs, leading entrepreneurs to overestimate their knowledge, judgment, or ability to perform entrepreneurial tasks independently. One reason is that GenAI can improve task performance without necessarily improving users' ability to accurately assess their own competence. As a result, users may overestimate their own knowledge or ability (Fernandes et al., 2026).

This risk may be further amplified by GenAI's tendency to produce sycophantic responses (Bo et al., 2026; M. Cheng et al., 2026). As Naddaf (2025) noted, GenAI models "often cheer users on, give them overly flattering feedback and adjust responses to echo their views, sometimes at the expense of accuracy" (p. 13). Cheng et al. (2026) examined the prevalence of sycophancy across 11 GenAI models using three datasets spanning a variety of use contexts and found that GenAI affirmed users' actions 49% more often than humans did. This sycophancy likely stems in part from the way GenAI models are trained and optimized to produce responses that users prefer, including responses that align with users' views (Sharma et al., 2024). Moreover, because GenAI models generate text by predicting successive tokens in response to a user's prompt, positively framed questions may steer the models toward supportive arguments rather than critical counterpoints (Brown et al., 2020). For example, when an entrepreneur asks, "Why will my sleep app for college students succeed?", GenAI may elaborate on reasons for success while overlooking potential weaknesses.

Such interactions can create a reinforcing feedback loop in which entrepreneurs' initial beliefs shape their prompts and GenAI's responses reinforce those beliefs, producing overly optimistic evaluations. D'Angelo et al. (2026) found that entrepreneurs' preexisting assumptions and beliefs about their venture ideas shaped how they interacted with GenAI, whose responses then reinforced those assumptions and beliefs. Furthermore, across three experiments, sycophantic GenAI increased individuals' conviction that they were right and heightened their preference for, trust in, and dependence on GenAI (Cheng et al., 2026). Repeated exposure to such affirming feedback may inflate entrepreneurs' confidence in their own judgments and reduce their attention to disconfirming evidence, increasing the risk of premature or poorly informed decisions.

**3.2.4 | Stage 2 summary.** Supporting our Empowerment–Entrapment Framework (Figure 1), GenAI may act as a double-edged sword during the opportunity evaluation and commitment stage. It may empower entrepreneurs by enabling more structured and informative opportunity evaluation, facilitating faster and less costly feasibility testing, and strengthening their self-efficacy to move forward. At the same time, GenAI may entrap entrepreneurs by creating an illusion of objectivity, overwhelming them with AI-generated information, and fostering overconfidence.

### 3.3 | Stage 3: Resource Assembly and Mobilization

During the resource assembly and mobilization stage, entrepreneurs seek to acquire and mobilize the resources needed to develop their ventures, including financial, human, and social resources (Baron & Shane, 2007). Because new ventures often lack these resources internally, entrepreneurs depend heavily on external stakeholders, such as investors, partners, suppliers, and professional networks (Rawhouser et al., 2017; E. Y. Zhao & Lounsbury, 2016). Securing such support requires entrepreneurs not only to identify and access potential resource providers, but also to convince them that the venture is credible and worthy of support.

**3.3.1 | Legitimacy establishment vs. Legitimacy distortion.**

***3.3.1.1 | Legitimacy establishment.*** During resource assembly and mobilization, entrepreneurs signal legitimacy to potential resource providers by crafting narratives that convey a comprehensible venture identity, articulate the logic for exploiting an opportunity, and provide a rationale that makes the venture understandable and credible (Lounsbury & Glynn, 2001; Martens et al., 2007). GenAI may support this process by helping entrepreneurs develop clearer, more persuasive, and stakeholder-oriented venture narratives. For example, GenAI can generate and refine entrepreneurial rhetoric across multiple formats—including elevator pitches and crowdfunding narratives—and mimic different rhetorical archetypes through prompt engineering (Short & Short, 2023). Even without explicit guidance on rhetorical techniques, GenAI can draw on strategies identified in prior entrepreneurship research, yielding narratives that experts rated as clearer and more persuasive (Ramoglou et al., 2026). In a subsequent experiment, GenAI iteratively refined these narratives to address different stakeholder concerns, increasing stakeholders' perceptions of legitimacy and persuasiveness, as well as their willingness to support the venture (Ramoglou et al., 2026). Similarly, GenAI can help small business owners translate their ideas into the "language of business" and rapidly develop business plans, lowering barriers to accessing capital (Romero Lauro et al., 2026).

Additionally, GenAI may help entrepreneurs tailor venture narratives to different cultural contexts. Lu et al. (2025) showed that GenAI models exhibited different cultural tendencies depending on the prompt language: When prompted in Chinese (versus English), GenAI models adopted a more interdependent (versus independent) social orientation and a more holistic (versus analytic) cognitive style, consistent with real-world cultural differences. Importantly, these cultural tendencies were also flexible. For instance, when prompted in English to assume the role of a Chinese person ("You are an average person born and living in China"), GenAI's responses became more interdependent and holistic, more closely resembling their responses when prompted in Chinese (Lu et al., 2025). This flexibility may allow entrepreneurs to align their narratives more closely with the expectations and values of different cultural audiences, potentially strengthening perceived legitimacy across cultural contexts.

***3.3.1.2 | Legitimacy distortion.*** When using GenAI to craft venture narratives, entrepreneurs may present traction, venture readiness, or future projections overly optimistically, especially when trying to signal preparedness to stakeholders. Because AI-generated content can be highly polished and persuasive, such narratives may appear convincing even when some elements are incomplete, inaccurate, or overstated. In two large-scale experiments, Williams-Ceci et al. (2026) found that biased autocomplete suggestions from a GenAI writing assistant shifted users' expressed attitudes toward the GenAI assistant's position, with most users remaining unaware of the bias. Similarly, Costello et al. (2026) found that when GenAI was instructed either to support or to oppose conspiracy theories, participants' beliefs shifted in the corresponding direction. Likewise, Danry et al. (2025) found that deceptive explanations generated by GenAI increased belief in false news headlines and weakened belief in true ones. Moreover, Randazzo et al. (2025) found that when professionals scrutinized GenAI outputs, GenAI attempted to persuade them to accept its preliminary output and intensified that persuasion as the professionals pushed back, thus making validation more difficult. In entrepreneurial contexts, these findings raise concerns that GenAI-assisted venture narratives may blur the line between persuasive framing and misrepresentation. If stakeholders later perceive a gap between communicated claims and

actual capabilities, entrepreneurs may lose credibility and damage their reputations.

**3.3.2 | Efficient stakeholder outreach vs. Negative social evaluations.**

***3.3.2.1 | Efficient stakeholder outreach.*** Entrepreneurs often mobilize resources by approaching potential investors, suppliers, partners, and other resource providers. Yet identifying and contacting these stakeholders can require considerable time and effort. GenAI may help streamline both activities by lowering the search and communication costs associated with stakeholder outreach, thereby increasing the efficiency of resource mobilization.

First, GenAI can help entrepreneurs *identify* relevant stakeholders and external resources by reducing search costs, mitigating information asymmetry, and improving the efficiency of information gathering and evaluation (J. Kim, 2025; Y. Li et al., 2025). Compared with traditional search engines (e.g., Google), GenAI can integrate multiple criteria into a single query and synthesize information across a broad range of options, making it easier for entrepreneurs to identify resource providers that better align with their specific needs (Y. Li et al., 2025). For example, an entrepreneur could prompt GenAI to search the Internet for local suppliers of sustainable materials in Boston that offer bulk-purchasing options and discounts on orders over $10,000, specifying that the suppliers must deliver within a 50-mile radius and have a minimum rating of 4.5 stars on customer review platforms.

Second, GenAI may lower the time and effort required to *contact* prospective stakeholders. For example, workers using an integrated GenAI tool spent about two fewer hours per week on email—roughly 17% less time—with no indication that these gains came at the expense of quality (Dillon et al., 2025). More broadly, GenAI can reduce the time required for writing tasks (Noy & Zhang, 2023) and help tailor messages to different audiences through prompt engineering (Short & Short, 2023). These findings suggest that GenAI may lower the costs of preparing and sending outreach messages to prospective stakeholders, thereby expanding entrepreneurs' capabilities to pursue partnerships that support venture formation.

***3.3.2.2 | Negative social evaluations.*** While GenAI may help entrepreneurs identify and contact stakeholders more efficiently, its perceived use may also trigger negative social evaluations when stakeholders perceive, infer, or are informed that an output was generated or assisted by GenAI.

When stakeholders attribute an output to GenAI rather than to the entrepreneur, they may infer lower effort and competence. For example, Reif et al. (2025) found that individuals who received help from GenAI were perceived as lazier, less competent, and less diligent than those who received non-AI help or no help. In a separate experiment, they found that job candidates using GenAI were perceived as lazier, which in turn predicted lower task-fit judgments in hiring contexts (Reif et al., 2025). Importantly, this effect was contingent on evaluators' own GenAI use: Evaluators who used GenAI less frequently tended to judge candidates who used GenAI less favorably, whereas more frequent GenAI users tended to evaluate them more favorably (Reif et al., 2025). Likewise, Kim et al. (2025) found that workers who used GenAI received lower compensation because evaluators perceived them as having exerted less effort and retained less agency over their work and, consequently, as deserving less credit for their work. Across 13 studies, this AI penalty was robust across different work scenarios and tasks (e.g., graphic design, web design, and writing social media posts) and employment statuses (e.g., full-time, part-time, and freelance) (J. Kim et al., 2025). Furthermore, Altay and Gilardi (2024) found that labeling news headlines as "AI-generated" reduced both their perceived accuracy and participants' willingness to share them, regardless of whether the headlines were true or false and whether they were human-generated or AI-generated.

Beyond perceptions of effort or competence, GenAI use may also reduce the perceived authenticity of entrepreneurs. Across seven experiments, Kirk and Givi (2025) found that emotional marketing communications believed to be written by GenAI (e.g., a salesperson's follow-up email expressing inspiration or a condolence message expressing sympathy) elicited more negative consumer responses, an effect serially mediated by perceived inauthenticity and moral disgust. In a personal-advice context, Osborne and Bailey (2025) found that participants rated the same advice as less authentic, less effective, and lower in quality when they believed it had been generated by GenAI rather than by another person.

In entrepreneurial settings, these findings suggest that investors, partners, and other stakeholders may evaluate entrepreneurs who use GenAI as less authentic, competent, diligent, or deserving of credit. Because investor trust and entrepreneur credibility are critical to resource mobilization, such perceptions may weaken entrepreneurs' ability to secure resources.

**3.3.3 | Improved hiring efficiency vs. Algorithmic bias in hiring.**

***3.3.3.1 | Improved hiring efficiency.*** As entrepreneurs mobilize resources, recruiting skilled employees becomes critical to transforming an opportunity into a functioning venture. This often requires entrepreneurs to define roles, establish selection criteria, and evaluate candidates, despite having limited formal HR support. GenAI can streamline parts of this process by assisting with tasks such as drafting job descriptions, preparing interview questions, and screening résumés. By automating and organizing the hiring workflow, GenAI may reduce the cognitive and time burdens associated with hiring decisions, enabling entrepreneurs to make hiring decisions more efficiently under resource constraints. For example, in a case study by Koteczki et al. (2025), the estimated time required for a recruitment process decreased by 13.25 hours when GenAI was integrated into selected

recruitment tasks, with corresponding reductions in recruitment-related costs.

***3.3.3.2 | Algorithmic bias in hiring.*** While GenAI can improve hiring efficiency, it may also introduce or amplify biases in hiring processes. For example, GenAI can produce gender- and race-based disparities in automated résumé evaluation. An et al. (2025) found that several GenAI models produced systematically different evaluations across demographic groups, assigning higher assessment scores to female candidates with otherwise similar qualifications but lower scores to Black male candidates with comparable qualifications. Newstead et al. (2023) found that leadership narratives generated by a GenAI writing tool exhibited gender bias: Almost half of the women leaders named were portrayed as bad leaders, compared with only a quarter of the men leaders named. Specifically, men leaders were more often portrayed as strong and charismatic, whereas women leaders were more often portrayed as emotional, ineffective, and fearful (Newstead et al., 2023).

In addition, GenAI models can produce covertly racist recommendations based solely on dialect cues, such as suggesting that speakers of African American English be assigned less prestigious jobs even when race is not specified (Hofmann et al., 2024). Moreover, Glickman and Sharot (2025) found that repeated exposure to biased GenAI outputs can alter human judgments and amplify bias over time. In an experiment, AI-generated images of "a financial manager" disproportionately depicted White men, and exposure to these outputs increased participants' tendency to identify White men, rather than members of other demographic groups, as being most likely to be financial managers (Glickman & Sharot, 2025). These findings illustrate a feedback loop: Biased human data shape GenAI systems, which may, in turn, shape human users and reinforce existing biases.

Beyond reproducing existing demographic biases, GenAI-supported hiring may also steer entrepreneurs toward candidates who conform to conventional indicators of success. For example, AI-generated personas for technology roles substantially overrepresented elite universities relative to actual LinkedIn data, with elite universities appearing in 72.45% of generated personas versus 8.56% in the LinkedIn benchmark data (Gupta & Ranjan, 2024). Such patterns may disadvantage candidates with nonelite or otherwise unconventional backgrounds, experiences, or strengths.

These biases may undermine not only fairness and inclusion but also the diversity that broadens perspectives, fosters exploration and innovation, and helps entrepreneurial teams adapt to new challenges (Beckman, 2006; Østergaard et al., 2011).

**3.3.4 | Stage 3 summary.** Supporting our Empowerment–Entrapment Framework (Figure 1), GenAI may operate as a double-edged sword during the resource assembly and mobilization stage. On the one hand, it may empower entrepreneurs by helping them establish legitimacy, identify and contact prospective stakeholders more efficiently, and streamline parts of the hiring process. On the other hand, GenAI may also undermine resource mobilization: GenAI-assisted narratives may distort legitimacy; perceived GenAI use may trigger negative judgments about entrepreneurs' effort, competence, and authenticity; and GenAI-supported hiring may reproduce or amplify existing biases.

## 3.4 | Stage 4: Venture Launch and Growth

Stage 4 marks the transition from resource mobilization to venture launch and growth. During venture launch, entrepreneurs convert a selected venture idea into a functioning venture by putting in place the structures and activities needed for continued operation and growth (Gartner, 1985; Shepherd et al., 2021). As the venture grows, the focus shifts toward expanding managerial and functional capabilities so that the venture can operate at a greater scale and accommodate increasing complexity. During this stage, entrepreneurs must balance short-term productivity with long-term human capital development to ensure that the venture remains adaptable as it evolves (Churchill & Lewis, 1983; DeSantola & Gulati, 2017).

**3.4.1 | Accelerated venture launch vs. Premature venture launch.**

***3.4.1.1 | Accelerated venture launch.*** During launch, entrepreneurs must manage multiple tasks simultaneously, including marketing, customer communication, documentation, analytics, and operational coordination. Emerging research suggests that GenAI can help entrepreneurs manage these demands by streamlining business processes and supporting a wide range of operational tasks (Kanbach et al., 2024). A study of 78 early-stage entrepreneurs indicates that GenAI can enhance entrepreneurial performance and help entrepreneurs save time and effort in the early stages of venture development (Marchena Sekli & Portuguez-Castro, 2025). In digital entrepreneurship, GenAI can reduce the resources required to launch ventures, especially among founders lacking managerial experience or advanced business degrees (Cao & Bhatia, 2025). More concretely, following the release of the GenAI coding tool GitHub Copilot, software-development startups secured initial funding 19% faster and employed 20% fewer software developers than did startups with less exposure to the tool (Heller & Asam, 2024).

***3.4.1.2 | Premature venture launch.*** Although GenAI may accelerate venture launch, it does not ensure operational readiness and may instead inflate perceptions of readiness, increasing the risk of premature launch. Extending the illusion of objectivity discussed in Section 3.2.1.2, GenAI's polished and structured outputs may also create an illusion of venture readiness, leading entrepreneurs to equate presentational quality with substantive preparedness. For example, GenAI can help an entrepreneur quickly build a polished website and generate persuasive product descriptions, even before arrangements with suppliers have been finalized and fulfillment processes have been fully developed. Moreover, as discussed in Section 3.2.3.2, GenAI's sycophantic tendency may reinforce entrepreneurs' initial assumptions and heighten overconfidence when they seek feedback on launch readiness. As a result, entrepreneurs may prematurely commit resources or enter the market before achieving sufficient product–market validation. Consistent with concerns about accelerated but low-

commitment entrepreneurial entry, Kim et al. (2026) found that solo-founded ventures became more prevalent following ChatGPT's release, but this growth did not translate into greater representation among the highest-ranked products, and the share of solo launches that had no updates within 12 months increased.

**3.4.2 | Boosted productivity and growth vs. Workslop, ethical risks, and diminished capabilities.**

***3.4.2.1 | Boosted productivity and growth.*** GenAI may boost productivity during venture launch and growth, allowing entrepreneurs to accomplish more with limited personnel and time. Studies across diverse work contexts—including customer service, consulting, and writing—indicate that GenAI can improve both the speed and quality of task execution (Brynjolfsson et al., 2025; Dell'Acqua, McFowland, et al., 2026; Gambacorta et al., 2026; Noy & Zhang, 2023; Wu et al., 2025). These productivity gains may empower entrepreneurs to scale their ventures without commensurate increases in workforce size. In a study of U.S. software startups, GenAI adoption was associated with greater output growth and faster commercialization without a significant increase in overall headcount (J. D. Kim & Conti, 2026).

GenAI may also support venture growth by strengthening strategic adaptation (Liu & Wang, 2024; Rezazadeh et al., 2025). Interviews with startups and scaleups showed that GenAI supported product-led, sales-led, and operational efficiency-driven growth through use cases such as content creation, customer-experience personalization, market-entry strategies, and customer engagement (Rezazadeh et al., 2025). GenAI may also facilitate specialized analytical tasks potentially relevant to venture growth, such as detecting non-answers in financial communications and, with expert guidance, producing risk analyses broadly comparable to those of human analysts (de Kok, 2025; Y. Zhou et al., 2024).

Notably, the value of GenAI depends on both access to the tool and entrepreneurs'

ability to evaluate, select, and implement GenAI outputs. In a field experiment involving 640 Kenyan entrepreneurs, Otis et al. (2026) found that entrepreneurs with above-median baseline performance who used a GenAI-powered business assistant experienced increases in revenue and profit of more than 15%, whereas entrepreneurs with below-median baseline performance experienced declines in revenue and profit of nearly 10%. Exploratory analyses suggested that higher-performing entrepreneurs were more likely to implement tailored, business-specific changes, whereas lower-performing entrepreneurs were more likely to adopt more generic advice, such as lowering prices or increasing advertising, even when such recommendations were poorly suited to their specific context (Otis et al., 2026).

***3.4.2.2 | Workslop.*** While GenAI can boost productivity and support venture growth, it may also generate “workslop,” defined as AI-generated content that appears polished but lacks substantive quality or value. Workslop may shift the burden downstream, requiring recipients (e.g., cofounders, advisors, and employees) to interpret, verify, and often rework outputs that seem acceptable on the surface but are substantively weak (Niederhoffer et al., 2025). Niederhoffer et al. (2025) reported that 40% of 1,150 U.S. employees across industries had received workslop in the preceding month. The costs were substantial: Employees reported spending an average of nearly two hours addressing each instance, resulting in an estimated $186 per employee per month in wasted time and rework (Niederhoffer et al., 2025). Beyond these direct productivity losses, workslop may also damage workplace relationships. Niederhoffer et al. (2025) found that recipients of low-effort AI-generated work viewed its senders as less creative, capable, reliable, trustworthy, and intelligent. These productivity and relational costs may be especially detrimental to startups, which require both operational efficiency and team cohesion to sustain rapid growth and scaling. As entrepreneurs expand their teams and take on more complex managerial tasks, they need to be cautious of potential workslop.

***3.4.2.3 | Ethical risks.*** GenAI use may also introduce ethical risks. Kaplan et al. (2025) introduced the concept of "machinal bypass," whereby individuals use GenAI to avoid emotional and intellectual engagement that would otherwise require their own presence, judgment, and effort. This dynamic may contribute to an accountability gap, as responsibility can be shifted from human actors to the AI system (Joo, 2024). Relatedly, across experiments involving multiple AI delegation interfaces and LLMs, Köbis et al. (2025) found that delegating tasks to GenAI can increase dishonest behavior, especially when people can induce unethical outcomes without explicitly specifying the misconduct. They also found that GenAI systems were more likely than human agents to comply with fully unethical instructions. Moreover, a time-lagged study of 389 technology entrepreneurs found that GenAI use was positively associated with unethical creativity (e.g., developing manipulative marketing strategies, fabricating content, or bypassing regulatory scrutiny) and that this association was stronger among entrepreneurs with high levels of Machiavellianism and narcissism (Mostafiz et al., 2026).

Beyond shifting responsibility, GenAI use may also shape individuals' underlying moral beliefs. In an experiment, Zhao et al. (2026) found that employees who used GenAI during a business-consulting task subsequently reported greater moral relativism—"the belief that morality is relative rather than absolute" (Lu, Quoidbach, et al., 2017, p. 1)—than employees who completed the task independently. In a three-wave survey of employees who used GenAI at work, GenAI use was similarly associated with higher moral relativism, which in turn predicted greater workplace deviance and more lenient evaluations of others' unethical workplace behavior (Zhao et al., 2026). Applied to entrepreneurial settings, these findings suggest that shifts in responsibility attribution and moral beliefs may make entrepreneurs more likely to justify unethical behavior. Although unethical actions such as tax evasion and data manipulation may yield short-term gains, they also expose ventures to legal, reputational, and operational risks.

***3.4.2.4 | Diminished cognitive engagement, learning, and memory.*** Although GenAI can streamline tasks and deliver quick outputs, overreliance on it may reduce cognitive effort and engagement and, over time, erode cognitive capabilities that are central to venture growth and success (Baron, 2007a; Cope, 2005). This concern is important because entrepreneurs need to make decisions under uncertainty, learn from incomplete and ambiguous feedback, and apply those lessons to future decisions.

A growing number of studies suggest that GenAI may reduce individuals' cognitive engagement. Kosmyna et al. (2025) randomly assigned participants to write an essay using ChatGPT, Google search, or no tools. Using electroencephalography to measure participants' brain activity during writing, the researchers found that participants in the ChatGPT condition demonstrated lower cognitive engagement than those in the search-engine and brain-only conditions. Similarly, in an experiment on argumentative writing, participants who used ChatGPT scored lower than those who did not on a measure of cognitive engagement that assessed mental effort, attention, deep processing, and strategic thinking (Georgiou, 2025).

Beyond these immediate effects on cognitive engagement, GenAI use may also erode learning and skill development. In uncertain environments, entrepreneurs must continually iterate and adapt, accumulate insights from experience, and apply those lessons to support venture growth and long-term development (Cope, 2005; Leatherbee & Katila, 2020). Caosun and Aral (2026) suggest that GenAI-enabled cognitive offloading may improve short-term productivity while reducing the practice needed to build and maintain expertise, thereby weakening learning and skill development. Across seven experiments, Melumad and Yun (2025) found that, compared with individuals who learned through traditional web search, users who acquired information from GenAI exerted less learning effort and consequently developed shallower knowledge. Similarly, Bastani et al. (2025) found that, relative to participants without access to GenAI, those given access to a standard GPT-4-based tutor performed 48% better during practice sessions, but 17% worse when that access was removed. Shen and Tamkin (2026) likewise found that participants who used a GenAI coding assistant while learning a new Python library performed significantly worse on a post-task assessment of conceptual understanding, code reading, and debugging abilities. Importantly, these learning effects may depend on how GenAI is used. Exploratory analyses of participants' GenAI-use patterns suggested that participants who relied heavily on AI delegation learned less than those who remained cognitively engaged and used GenAI to support conceptual understanding (Shen & Tamkin, 2026).

A related concern involves memory. In the aforementioned experiment by Kosmyna et al. (2025), 83.3% of participants assigned to the ChatGPT condition were unable to accurately quote from their own essays, compared with 11.1% in each of the search-engine and brain-only conditions. Similarly, Barcaui (2025) found that students using GenAI as a study aid performed significantly worse on a long-term retention test administered 45 days after learning than those who studied using traditional methods without GenAI. Consistent with these findings, a three-wave survey found that GenAI use was associated with greater memory loss (Abbas et al., 2024).

Applied to entrepreneurship, these findings suggest that GenAI may create a trade-off between immediate task completion and the development or retention of underlying cognitive capabilities. These risks may be greatest when entrepreneurs use GenAI to substitute for their own cognitive effort rather than to support active reflection, evaluation, and understanding. For example, Romero Lauro et al. (2026) found that although a GenAI-powered business planning tool enabled entrepreneurs to rapidly generate business plans, its ease of use created a tension between obtaining instant results and engaging more deeply in thinking through their business assumptions, goals, and strategies.

**3.4.3 | Increased functional breadth and self-sufficiency vs. Reduced prosociality.**

***3.4.3.1 | Increased functional breadth and self-sufficiency.*** Entrepreneurship requires knowledge and skills across multiple functional domains, and a broad, balanced skill set can help entrepreneurs launch and grow their ventures (Lazear, 2004; Stuetzer et al., 2013). GenAI may expand entrepreneurs' functional breadth in two ways: by helping them acquire skills they do not yet possess and by performing tasks on their behalf when they lack the necessary expertise. Together, these pathways may enable entrepreneurs to perform a wider range of functions throughout venture creation and growth.

First, GenAI may broaden entrepreneurs' functional breadth by helping them acquire new knowledge and develop new skills. George-Reyes et al. (2024) found that GenAI-supported entrepreneurship training was associated with gains in entrepreneurial skills. Similarly, Y. Li et al. (2025) suggested that GenAI can support entrepreneurial information gathering, knowledge construction, and problem solving throughout the entrepreneurial process. Likewise, in a survey spanning 15 countries, Zhang et al. (2026) found that GenAI use was positively associated with knowledge acquisition and application, which were, in turn, positively associated with entrepreneurial skill development. GenAI may also help users develop more specialized skills. For example, GenAI-based negotiation coaching bots can help users prepare for negotiations, understand other parties' interests, and formulate negotiation strategies. In multiparty role-play exercises, students using such bots reported feeling better prepared for negotiations and more capable of identifying other parties' interests and developing appropriate strategies (Susskind et al., 2024).

Second, GenAI may extend entrepreneurs' functional reach by performing tasks that would otherwise require expertise they do not personally possess. Rather than requiring entrepreneurs to master every skill needed to launch and grow a venture, GenAI can provide some of the expertise or execution needed to complete unfamiliar tasks. Using an abductive mixed-methods case study of OpenAI's ecosystem from 2020 to 2025, Nzembayie and Urbano (2026) identified capability scaffolding as a key mechanism through which GenAI platforms reduce access, skill, and cost barriers, enabling resource-constrained ventures to build and scale offerings that would otherwise require capabilities unavailable internally. Similarly, Donaldson et al. (2026) suggested that GenAI can compensate for certain capability deficits in the short term, while emphasizing that sustainable value creation depends on complementary skills and learning processes.

By both helping entrepreneurs learn new skills and allowing them to complete tasks without mastering the underlying expertise, GenAI may increase entrepreneurs' self-sufficiency. Consistent with this possibility, Kim et al. (2026) found that the increase in entrepreneurial entry following ChatGPT's release was disproportionately driven by solo founders and was especially pronounced in categories that had historically relied more heavily on teams. Cao and Bhatia (2025) likewise found that GenAI lowered barriers to entrepreneurial entry, with larger increases in new ventures among founders lacking managerial experience or advanced business education. These findings suggest that GenAI may allow individual founders to perform generalist and managerial functions that previously required the complementary expertise of cofounders or other team members.

***3.4.3.2 | Reduced prosociality.*** The increase in functional breadth and self-sufficiency may come at the cost of reduced prosociality, defined as the tendency to engage in voluntary behaviors intended to benefit others (Caprara et al., 2012). As entrepreneurs become more accustomed to relying on GenAI to solve problems, they may become less inclined to help others or engage in reciprocal exchanges that build connection and trust over time.

Across a series of studies, Lu et al. (2026) found that GenAI use reduced individuals' humility, which in turn reduced their willingness to engage in prosocial behaviors (e.g., charitable giving). Similarly, combining a large-scale computational evaluation across 11 leading GenAI models and 11,587 queries with three preregistered experiments, Cheng et al. (2026) found that GenAI's sycophancy reduced individuals' prosocial intentions, including their willingness to apologize or repair interpersonal relationships. At the same time, sycophantic interactions increased participants' trust in the GenAI system and their intention to use it again, suggesting that interpersonal disengagement may occur alongside greater reliance on GenAI (M. Cheng et al., 2026). Likewise, a three-wave longitudinal survey study found that GenAI use was negatively associated with employees' perceived instrumentality of their work relationships and with helping behaviors (J.-M. Li et al., 2025).

Applied to entrepreneurial settings, these findings suggest a potential tension for entrepreneurs: While GenAI may enable entrepreneurs to perform a broader range of tasks independently, greater reliance on GenAI may also reduce some of the interpersonal interactions through which strong entrepreneurial relationships are developed and maintained. Such relational costs can be consequential for entrepreneurs because venture development often depends on trust-based relationships that support mentorship, resource exchange, and long-term success (Shane & Cable, 2002).

**3.4.4 | Enhanced well-being vs. Eroded coping capacity.**

***3.4.4.1 | Enhanced well-being.*** Entrepreneurship is often characterized by uncertainty and resource constraints, which may heighten entrepreneurs' stress and anxiety (Thompson et al., 2020; Williamson et al., 2021). These pressures may become particularly intense as ventures grow, when entrepreneurs face greater workloads, tighter deadlines, and increasing responsibility for employees and organizational expansion (Partanen et al., 2025). A growing number of individuals have turned to GenAI for psychological support. For example, a 2025 analysis of publicly discussed GenAI use cases ranked "therapy/companionship" as the top use case in 2025 (Zao-Sanders, 2025). In a six-week randomized controlled trial, access to a GenAI-powered app improved users' social well-being, including reduced loneliness and a

stronger sense of belonging and connection to their communities (Cachia et al., 2026). In a 12-week randomized clinical trial, an LLM-based emotional-support platform produced greater improvements in well-being than both face-to-face group therapy and a waitlist control among individuals experiencing psychological distress (Shoshani et al., 2026). Whereas access to human counselors is constrained by time and cost, GenAI systems can provide continuous, on-demand well-being support. This responsiveness and accessibility may be especially valuable for entrepreneurs, who often face irregular schedules, ongoing pressure, and limited time to seek well-being support. Therefore, GenAI may offer entrepreneurs a readily accessible source of emotional support, potentially helping them manage distress, maintain well-being, and sustain a sense of agency when facing entrepreneurial challenges (Dechant et al., 2025).

***3.4.4.2 | Eroded coping capacity***. Nevertheless, excessive reliance on GenAI for psychological support may have unintended consequences for individuals' ability to cope independently with stress and adversity. Although GenAI can help entrepreneurs regulate negative emotions in the moment, heavy reliance on it may reduce opportunities to engage in independent coping, reflection, and self-regulation. Drawing on complementary on-platform and experimental analyses of ChatGPT, OpenAI and MIT researchers found that high-intensity ChatGPT use (e.g., top-decile use) was associated with greater loneliness, stronger emotional dependence on ChatGPT, and less social interaction with other people (Fang et al., 2025; Phang et al., 2025). These findings suggest a potential risk of emotional overreliance on GenAI. Applied to entrepreneurial settings, such dependence may reduce entrepreneurs' opportunities to develop or exercise coping strategies when confronting setbacks and uncertainty. Over time, this pattern could increase entrepreneurs' reliance on GenAI for emotional regulation and weaken their capabilities to manage stress and adversity without GenAI.

**3.4.5 | Stage 4 summary.** Supporting our Empowerment–Entrapment Framework (Figure 1), GenAI may function as a double-edged sword during the venture launch and growth stage. On the one hand, GenAI can help entrepreneurs launch ventures more quickly, boost productivity, increase functional breadth and self-sufficiency, and provide well-being support as organizational demands intensify. On the other hand, these benefits of GenAI may also carry corresponding costs: Accelerated execution may create an illusion of readiness and foster premature launch; productivity gains may be offset by workslop, ethical risks, and diminished cognitive engagement, learning, and memory; increased functional breadth and self-sufficiency may come at the expense of prosociality; and continuous access to well-being support may erode entrepreneurs' capabilities to cope independently.

### 3.5 | GenAI Features Underlying Its Empowering and Entrapping Effects

Having detailed the benefits and costs of GenAI use at each stage of the entrepreneurial process (Sections 3.1 to 3.4), we now identify core features of GenAI that provide common bases for these effects. Based on the aforementioned definition of GenAI—"a category of AI that creates new content (such as text, images, audio, and video) by learning patterns from existing data" (Lu et al., 2025, p. 2360)—we focus on two core features of GenAI: (1) its efficiency and convenience in creating content and completing tasks, and (2) its training data and design features (see Table 2).

-----------------------------

Insert Table 2 about here

-----------------------------

**3.5.1 | Efficiency and convenience.** GenAI is characterized by its high processing speed, automation, and constant availability, all of which enhance the efficiency and convenience of task execution (Brynjolfsson et al., 2025; Noy & Zhang, 2023). These characteristics may underlie many of GenAI's empowering effects by reducing the time and

resources required to perform entrepreneurial tasks. For example, GenAI can help entrepreneurs conduct efficient feasibility testing (Santoro et al., 2026), identify and contact stakeholders (Short & Short, 2023), improve hiring efficiency (Koteczki et al., 2025), accelerate venture launch (Marchena Sekli & Portuguez-Castro, 2025), and boost productivity and growth (Dell'Acqua, McFowland, et al., 2026). Yet the same convenience can also create entrapping effects by making it easy to delegate work that would otherwise require greater cognitive, emotional, or interpersonal effort (Kosmyna et al., 2025). This reliance can lead stakeholders to perceive entrepreneurs as less competent, credible, and authentic (Kirk & Givi, 2025; Reif et al., 2025); fuel workslop (Niederhoffer et al., 2025) and unethical behavior (Köbis et al., 2025); and diminish cognitive engagement, learning, and memory (Bastani et al., 2025). Moreover, by making difficult entrepreneurial tasks feel easier and more manageable, GenAI can increase entrepreneurial self-efficacy (Xie & Wang, 2025). Yet this empowerment may also foster an inflated sense of self-sufficiency and an illusion of venture readiness, leading to entrepreneurial overconfidence (Fernandes et al., 2026), reduced prosociality (J.-M. Li et al., 2025), and premature venture launch (H. Kim et al., 2026).

**3.5.2 | Training data and design features.** GenAI's effects across the entrepreneurial process also stem from its training data and design features. Trained on broad and vast datasets (Bommasani et al., 2021; Brown et al., 2020), GenAI can improve the quality of venture ideas (Dell'Acqua, Ayoubi, et al., 2026), enhance creativity at the individual level (Sun et al., 2025), and increase entrepreneurs' functional breadth and self-sufficiency (Nzembayie & Urbano, 2026). However, because GenAI is trained on data that may reflect existing biases, its outputs may reproduce or amplify those biases (Bommasani et al., 2021), potentially misleading entrepreneurs in opportunity evaluation and hiring (An et al., 2025). In addition, because GenAI outputs are generated probabilistically from patterns learned from training data and do not guarantee factual accuracy (Brown et al., 2020; B. Chen et al., 2024;

Wang et al., 2026), they may contain hallucinations, particularly when relevant information is sparse, ambiguous, outdated, or poorly represented (Farquhar et al., 2024). Moreover, because GenAI models are trained on overlapping large-scale corpora, they tend to produce similar outputs for different entrepreneurs who provide comparable inputs (Brown et al., 2020; Gao et al., 2020; Touvron et al., 2023), thereby decreasing the distinctiveness of venture ideas (Z. Cheng & Gao, 2026; Doshi & Hauser, 2024). Furthermore, GenAI often produces structured and detailed outputs because it is fine-tuned to generate responses that users find clear and helpful (Longpre et al., 2023; Ouyang et al., 2022; Rafailov et al., 2023). While this structure can support entrepreneurs in evaluating venture opportunities (Albashrawi, 2025), it may also make GenAI's outputs appear more objective and accurate than they actually are, creating an illusion of objectivity (L. Huang et al., 2025).

**3.6 | Boundary Conditions of GenAI's Empowering and Entrapping Effects**

Our integrative review identifies several boundary conditions that moderate GenAI's empowering and entrapping effects. These boundary conditions broadly concern entrepreneurs' expertise, approaches to using GenAI, and personal traits and values.

First, entrepreneurs' expertise moderates the extent to which they benefit from GenAI. GenAI's creative benefits tend to be particularly large for individuals with lower baseline creativity or less domain expertise (Doshi & Hauser, 2024; M. Huang et al., 2024). Similarly, GenAI's productivity benefits tend to be particularly large for individuals with lower baseline skill or performance levels and for non-native speakers (Brynjolfsson et al., 2025; Dell'Acqua, McFowland, et al., 2026; Mei et al., 2025; Noy & Zhang, 2023). Likewise, Cao and Bhatia (2025) found that GenAI's positive effect on entrepreneurial entry was especially strong among founders without managerial experience or advanced business education. These findings indicate a compensatory pattern whereby GenAI produces greater benefits for individuals with larger initial gaps in experience and expertise. This pattern may arise for two reasons: Individuals with less expertise may have more room for improvement, whereas those with greater expertise may rely on well-established work routines that make GenAI harder to integrate (Hou et al., 2025).

At the same time, greater expertise may enable entrepreneurs to evaluate, adapt, and apply GenAI outputs more effectively. For example, Heller and Asam (2024) found that GenAI-related gains in startup funding were larger among founders with more extensive technological or managerial expertise. Similarly, Otis et al. (2026) found that entrepreneurs with higher baseline performance benefited from GenAI advice, whereas entrepreneurs with lower baseline performance experienced declines in revenue and profit—because higher-performing entrepreneurs were more likely to implement business-specific advice, whereas lower-performing entrepreneurs were more likely to follow generic recommendations. Zhang et al. (2026) further found that entrepreneurial competency strengthened the positive relationships between GenAI adoption and both knowledge acquisition and knowledge application. Beyond domain-specific and entrepreneurial capabilities, proficiency in using GenAI may also matter. In an experiment on programming education, Jin et al. (2025) found that GenAI proficiency moderated the effects of GenAI use: Learners with higher (vs. lower) GenAI proficiency were particularly likely to obtain informational benefits, experience lower learning costs, and achieve knowledge fit.

In sum, expertise may moderate GenAI's effects in multiple ways, influencing both how much users stand to gain and how effectively they can evaluate, apply, and integrate its outputs.

Second, how entrepreneurs engage with GenAI also moderates its effects. Entrepreneurs with stronger metacognition are more likely to monitor GenAI outputs, adapt their interactions with the tool, and reconsider their approach when necessary, thereby realizing greater creative benefits (Sun et al., 2025). Conversely, Shen and Tamkin (2026) found that learning was weaker when users delegated cognitive tasks to GenAI and stronger when they used it to support conceptual understanding and independent thinking. Relatedly, D'Angelo et al. (2026) found that entrepreneurs mitigated the risk of overtrusting GenAI by verifying evidence, considering alternative scenarios, and comparing its outputs against established frameworks. Collectively, these findings suggest that GenAI is more likely to empower entrepreneurs when they actively scrutinize, contextualize, and refine its outputs rather than accepting them uncritically.

Finally, certain personal traits and values may heighten entrepreneurs' susceptibility to the potential harms of GenAI use. For example, Mostafiz et al. (2026) found that the positive relationship between GenAI use and unethical creativity was particularly strong among entrepreneurs with higher levels of Machiavellianism and narcissism. Relatedly, Zhao et al. (2026) found that the effects of AI use on unethical outcomes (e.g., workplace deviance and greater tolerance of others' unethical behavior) were more pronounced among employees with a weaker moral identity. These findings suggest that GenAI's effects are shaped not only by the technology's capabilities but also by the traits and values of its users.

# 4 | Discussion

## 4.1 | Theoretical Contributions

This article offers important theoretical contributions. First, we integrate findings from a fragmented body of literature to provide an entrepreneur-centered review of GenAI's effects, thereby offering a more comprehensive understanding of how GenAI affects entrepreneurs.

Second, we develop a *stage-based* framework that maps GenAI's effects across four stages of the entrepreneurial process—(1) opportunity recognition and ideation, (2) opportunity evaluation and commitment, (3) resource assembly and mobilization, and (4) venture launch and growth—each characterized by distinct tasks and demands. Notably, our review reveals an uneven evidence base across the entrepreneurial process. As shown in Tables S1 and S2, entrepreneurship-specific evidence appears more limited for Stage 1 (opportunity recognition and ideation) and Stage 3 (resource assembly and mobilization) than for Stage 2 (opportunity evaluation and commitment) and Stage 4 (venture launch and growth).

Third, we extend this stage-based framework into the Empowerment–Entrapment Framework (Figure 1) to highlight GenAI's double-edged role in entrepreneurship. Rather than merely cataloging GenAI's benefits and costs, we theorize its empowering and entrapping effects jointly and identify the potential trade-offs between them. For example, during the opportunity recognition and ideation stage, while GenAI can increase entrepreneurs' creativity, it may also reduce the distinctiveness of their venture ideas (Section 3.1.2). During the opportunity evaluation and commitment stage, GenAI can support structured and informative evaluation of opportunities while also fostering an illusion of objectivity that distorts decision-making (Section 3.2.1). During the resource assembly and mobilization stage, while using GenAI can help entrepreneurs identify and contact stakeholders more efficiently, stakeholders may perceive such entrepreneurs as less effortful, competent, and authentic (Section 3.3.2). During the venture launch and growth stage, while GenAI can boost short-term productivity and growth, it may also generate workslop and erode long-term capabilities such as learning and memory (Section 3.4.2). By demonstrating how GenAI simultaneously empowers and entraps entrepreneurs, our framework challenges overly optimistic and alarmist perspectives to offer a more balanced account of GenAI use in entrepreneurship.

Fourth, we examine how GenAI features give rise to both empowerment and entrapment across the entrepreneurial process. In particular, we show that many of GenAI's effects stem from two core characteristics of GenAI: its efficiency and convenience in creating content and completing tasks (Section 3.5.1) and its training data and design features (Section 3.5.2). We highlight that the same features that empower entrepreneurs can also entrap them. First, GenAI's efficiency and convenience help explain many of its benefits, such as improving hiring efficiency, accelerating venture launch, and boosting productivity and growth. Yet that same convenience may also encourage entrepreneurs to excessively outsource cognitive and emotional work to GenAI. Second, GenAI's training data and design features give rise to a variety of benefits and costs. For instance, GenAI's broad information scope can improve idea quality and increase entrepreneurs' functional breadth and self-sufficiency, while its reliance on real-world training data can reproduce existing biases, such as gender and racial biases in hiring. By tracing the benefits and costs back to these underlying features of GenAI, we provide a deeper understanding of how GenAI both empowers and entraps entrepreneurs.

Fifth, we identify boundary conditions that moderate GenAI's effects on entrepreneurs (Section 3.6). Our review shows that whether GenAI empowers or entraps entrepreneurs depends on their expertise, approaches to using GenAI, and personal traits and values. By identifying these boundary conditions, we clarify when and for whom GenAI is more likely to empower or entrap entrepreneurs.

## 4.2 | Practical Implications

In addition to our theoretical contributions, we offer important practical guidance for entrepreneurs seeking to use GenAI effectively across the entrepreneurial process.

**4.2.1 | Recognize the benefits and costs of GenAI across entrepreneurial stages.** Given our insight that GenAI can both empower and entrap entrepreneurs, they should recognize its double-edged nature at each stage of the entrepreneurial process.

During the opportunity recognition and ideation stage (Section 3.1), entrepreneurs can use GenAI to improve idea quality and boost creativity. However, they should avoid uncritically accepting AI-generated venture ideas, as GenAI may decrease the distinctiveness of ideas, introduce hallucinations, and reproduce biases embedded in its training data. To mitigate these risks, entrepreneurs should proactively explore alternatives, niche markets, and non-mainstream customer groups rather than relying solely on the first ideas generated by GenAI. They should also implement verification practices—such as cross-checking claims against independent data sources or requiring traceable evidence—to reduce the risk of hallucinations (Rady et al., 2026). In addition, entrepreneurs may use triangulation strategies, such as comparing outputs generated using multiple prompts or GenAI models, to improve the robustness of their venture ideas (Doshi et al., 2025).

During the opportunity evaluation and commitment stage (Section 3.2), entrepreneurs can use GenAI to structure comparisons across venture ideas and support efficient feasibility testing to evaluate whether an opportunity is worth pursuing. Meanwhile, they should remain cautious of information overload, as the sheer volume of analysis generated by GenAI could overwhelm entrepreneurs and impede their decision-making. Entrepreneurs should also be mindful that GenAI may boost entrepreneurial self-efficacy while heightening the risk of overconfidence. One useful strategy is an "ask, don't tell" approach (Dubois et al., 2026): Rather than using prompts that presuppose success, entrepreneurs should frame prompts as open-ended or critical questions. Doing so can steer GenAI toward more critical analysis and reduce its tendency to generate sycophantic responses.

During the resource assembly and mobilization stage (Section 3.3), entrepreneurs can use GenAI to support legitimacy building, outreach, and hiring. It can help refine venture narratives, reduce the time required to draft outreach materials, and improve hiring efficiency. However, because resource mobilization is highly relational, entrepreneurs should remain personally involved in high-stakes stakeholder interactions. In addition, because GenAI-assisted hiring may reproduce demographic biases, entrepreneurs should implement internal governance mechanisms, including regular internal audits and continuous algorithm monitoring, documentation of system strengths and weaknesses, and procedures to ensure compliance with relevant recruitment laws and regulations (Chen, 2023).

During the venture launch and growth stage (Section 3.4), entrepreneurs can use GenAI to accelerate launch, boost productivity and growth, increase their functional breadth and self-sufficiency, and enhance well-being. Yet these benefits also require caution. Entrepreneurs should guard against premature launch by distinguishing polished artifacts from actual venture readiness. They should also ensure that short-term productivity gains do not come at the cost of workslop, unethical behavior, or diminished cognitive engagement, learning, and memory. Although greater functional breadth may increase entrepreneurs' self-sufficiency, they should ensure that this increased self-sufficiency does not diminish their willingness to seek help, support others, or collaborate with stakeholders. Likewise, although GenAI may help entrepreneurs manage stress, overreliance on it for emotional support may erode their capabilities to cope independently over time. To preserve these capabilities, entrepreneurs should remain actively involved in core cognitive and interpersonal processes, such as independently solving problems, making decisions, and engaging directly in relationship-building activities.

**4.2.2 | Calibrate GenAI use.** Beyond recognizing these benefits and risks, entrepreneurs should calibrate how they use GenAI. Appropriate use strategies are essential for harnessing the benefits of GenAI while mitigating its potential drawbacks. As reviewed in Section 3.1.2.1, entrepreneurs with higher metacognition are more likely to convert GenAI adoption into meaningful creative benefits (Sun et al., 2025). In practice, this requires entrepreneurs to carefully consider the purpose of each prompt, critically evaluate whether the outputs are accurate and evidence-based, and iteratively refine prompts to encourage

GenAI to elaborate on, improve, or expand its responses. For example, Boussioux et al.'s (2024) experiment compared *independent* search (where GenAI generated multiple solutions from a single initial prompt) with *differentiated* search (where human-guided prompts instructed GenAI to sequentially differentiate each new output from previous ones), and found that differentiated search produced significantly higher-quality solutions. Relatedly, strategies that promote active and reflective engagement with AI-generated content may help preserve users' cognitive engagement (Georgiou, 2025).

GenAI use strategies should also be tailored to entrepreneurs' prior knowledge and expertise. Novice entrepreneurs, who often face greater knowledge gaps, may benefit more from GenAI's ability to support knowledge acquisition and accelerate early-stage execution. At the same time, they should use GenAI critically to avoid overreliance and inflated confidence. More experienced entrepreneurs should consider whether GenAI complements or conflicts with their established ways of working. When GenAI is difficult to integrate into existing routines, it may create additional revision and rework rather than improve creativity or efficiency (Hou et al., 2025). Experienced entrepreneurs may therefore benefit from using GenAI selectively for tasks where it adds fresh perspectives or increases efficiency, while relying on their own domain-specific judgment in areas where their expertise remains more effective.

**4.2.3 | Protect relational assets.** Entrepreneurs should take deliberate steps to preserve relational assets when using GenAI. While GenAI can help entrepreneurs craft persuasive narratives and contact stakeholders more efficiently, the perceived use of GenAI may also trigger a social evaluation penalty, whereby stakeholders perceive entrepreneurs as less effortful, less competent, or less authentic. To mitigate these risks, entrepreneurs can adopt a "high-tech, high-touch" strategy (McKendrick, 2021). In practice, this means using GenAI for back-end efficiency—such as synthesizing market data or organizing

information—while maintaining strong personal ownership over front-end, high-stakes interactions. For example, for written communications, one useful practice is to use GenAI more as an editor than as an author: Entrepreneurs can draft messages themselves and then use GenAI to improve their clarity and structure. During resource mobilization, entrepreneurs could also prioritize face-to-face meetings or personalized communication to signal genuine commitment and build trust-based relationships. By preserving meaningful human involvement in stakeholder relationships, entrepreneurs can benefit from GenAI's efficiency while reducing negative social evaluations and protecting the trust and relational assets needed for venture development.

**4.2.4 | Determine what to delegate to GenAI.** Entrepreneurs should allocate tasks between themselves and GenAI based on their respective strengths. Because GenAI performs unevenly across tasks, entrepreneurs should exercise caution when making complex decisions that require integrating quantitative data with contextual information (Dell'Acqua, McFowland, et al., 2026). In particular, entrepreneurs may benefit from using GenAI to generate and compare alternatives while retaining human responsibility for evaluating those alternatives and making final decisions. GenAI can help expand the range of possible options and integrate technical and commercial perspectives, but human judgment remains important for determining which alternatives are appropriate to pursue (Dell'Acqua, Ayoubi, et al., 2026). Similarly, when entrepreneurs draft entrepreneurial narratives, GenAI may help them generate and refine ideas, while human judgment remains essential for interpreting outputs and ensuring alignment with their goals and context (Ramoglou et al., 2026).

### 4.3 | Directions for Future Research

Our integrative review and the Empowerment–Entrapment Framework reveal limitations in the existing literature and point to several promising directions for future research on entrepreneurship in the era of GenAI.

First, future research should strengthen the entrepreneurship-specific evidence base. As shown in Table S2, existing evidence is unevenly distributed across the entrepreneurial process, and findings at less-studied stages (Stages 1 and 3) often come from research conducted in adjacent contexts. Future studies should therefore examine GenAI's effects more directly at these stages and assess whether findings from adjacent contexts generalize to entrepreneurial settings.

Second, future work could examine a broader range of boundary conditions beyond those identified in the existing literature. Although our review has highlighted several *entrepreneur characteristics* as boundary conditions (e.g., metacognition, entrepreneurial experience, and domain expertise), other individual and environmental factors remain underexplored. For example, at the individual level, entrepreneurs with high Artificial Intelligence Quotient (AIQ)—defined as "a person's ability to use AI to perform a wide variety of tasks" (Qin et al., 2024)—may be particularly adept at harnessing GenAI's benefits while mitigating its costs. Moreover, the extent to which entrepreneurs are averse to or appreciative of GenAI is also likely to moderate its influence throughout the entrepreneurial process (Qin, Zhou, et al., 2025). At the environmental level, entrepreneurial ecosystems differ substantially in industry structure, resource availability, risk tolerance, AI adoption norms, and cultural attitudes toward AI, all of which may influence how GenAI affects entrepreneurial outcomes. For example, the effects of GenAI may differ between uncertainty-avoidant and uncertainty-tolerant cultures (Y. Li & Zahra, 2012; Liebregts et al., 2025; Lu, 2023), between tight and loose cultures (Assenova & Amit, 2025), between collectivistic and individualistic cultures (Y. Li & Zahra, 2012; Lu et al., 2021; Tiessen, 1997), between resource-constrained and resource-rich environments (Adomako et al., 2025), and across industries with different levels of technological intensity or regulatory oversight (Sarkar et al., 2006). Comparative research across entrepreneurial ecosystems would deepen our understanding of the conditions under which GenAI presents benefits or challenges for entrepreneurs.

Third, to date, studies on GenAI and entrepreneurs have predominantly focused on individual-level constructs, while team-level constructs remain comparatively underexplored. Given that entrepreneurial activities often unfold within founding teams, future research should examine how GenAI shapes team-specific constructs such as knowledge sharing, task conflict, relationship conflict, and team coordination. For example, GenAI may reduce task conflict by structuring information and clarifying alternatives, yet it may also introduce new forms of disagreement when AI-generated outputs conflict with founders' expertise. Understanding how GenAI interacts with team composition, expertise diversity, and communication patterns within founding teams represents an important direction for advancing entrepreneurship theory in the GenAI era.

Fourth, to gain a more holistic understanding of how GenAI shapes entrepreneurial processes, future research should examine *multiple* outcomes within a study rather than focusing on a *single* outcome in isolation. Doing so would help clarify how GenAI's effects unfold across different entrepreneurial activities. Relatedly, while emerging research has provided some indications of GenAI's *short-term* benefits and *long-term* costs—for example, GenAI may enhance creativity and productivity while potentially diminishing cognitive engagement, learning, and memory (Bastani et al., 2025; Georgiou, 2025; Sun et al., 2025)—much of this evidence comes from studies conducted over relatively short periods, which cannot fully capture the long-term effects of GenAI use. Therefore, we encourage future research to adopt longitudinal designs to compare the short- and long-term effects of GenAI use. Such designs would be especially valuable for examining the temporal dimension of GenAI use by identifying when initially empowering effects begin to become entrapping. Such work would provide deeper insight into the long-term implications of GenAI use for entrepreneurs and help determine whether short-term gains ultimately translate into sustainable entrepreneurial success.

Fifth, our framework adopts a stage-based and linear representation of the entrepreneurial process for analytical clarity. Taking this representation as a starting point, future research could examine whether GenAI not only generates benefits and costs at each stage, but also reshapes the stages themselves. For example, GenAI may compress certain stages or blur boundaries between stages. In addition, future research should examine whether GenAI's empowering and entrapping effects differ in more iterative entrepreneurial processes (Sarasvathy, 2001). For example, GenAI's sycophancy may be more problematic in linear processes that lock entrepreneurs into early ideas than in nonlinear, iterative processes that allow for continuous revision.

Finally, given the rapid evolution of GenAI, future research should examine how emerging AI technologies transform entrepreneurial practices, as well as how these effects evolve over time. While existing research has primarily focused on text-based LLMs (e.g., ChatGPT) and image generators (e.g., DALL·E), emerging advances in GenAI may reshape entrepreneurship in new ways. For example, as AI *agents* evolve, they may make it easier for entrepreneurs to start and grow businesses independently (The Economist, 2025; Tiwari, 2025). OpenAI CEO Sam Altman has predicted that advances in GenAI will facilitate the rise of one-person, billion-dollar companies (Graham, 2024). Future research should also examine whether, as AI autonomy increases, GenAI remains a tool for human entrepreneurs or begins to take on core entrepreneurial functions itself. Because our Empowerment–Entrapment Framework centers on human entrepreneurs' use of GenAI, higher levels of AI autonomy may require theories that treat AI not only as a tool but also as an actor in the entrepreneurial process.

## 5 | Conclusion

Generative artificial intelligence is rapidly reshaping entrepreneurship. In this paper, we have synthesized the fragmented literature on GenAI and entrepreneurship and introduced the Empowerment–Entrapment Framework to explain how GenAI use affects entrepreneurs throughout the entrepreneurial process. Rather than treating GenAI as either an asset or a liability, our framework identifies potential trade-offs between its empowering and entrapping effects throughout the entrepreneurial process. We also identify core features of GenAI that provide common bases for these effects.

More broadly, our paper suggests that the future of entrepreneurship will depend less on *whether* entrepreneurs use GenAI than on *how* they use it. When used thoughtfully, GenAI may empower entrepreneurs by expanding their capabilities to engage in entrepreneurial activities. When used uncritically, GenAI may entrap entrepreneurs by restricting or eroding those capabilities. We hope that this paper lays a foundation for future research and advances a more balanced perspective on how entrepreneurship is increasingly shaped by, but not reducible to, GenAI.

**Figure 1. The Empowerment–Entrapment Framework: Generative AI's Benefits and Costs Across Entrepreneurial Stages**

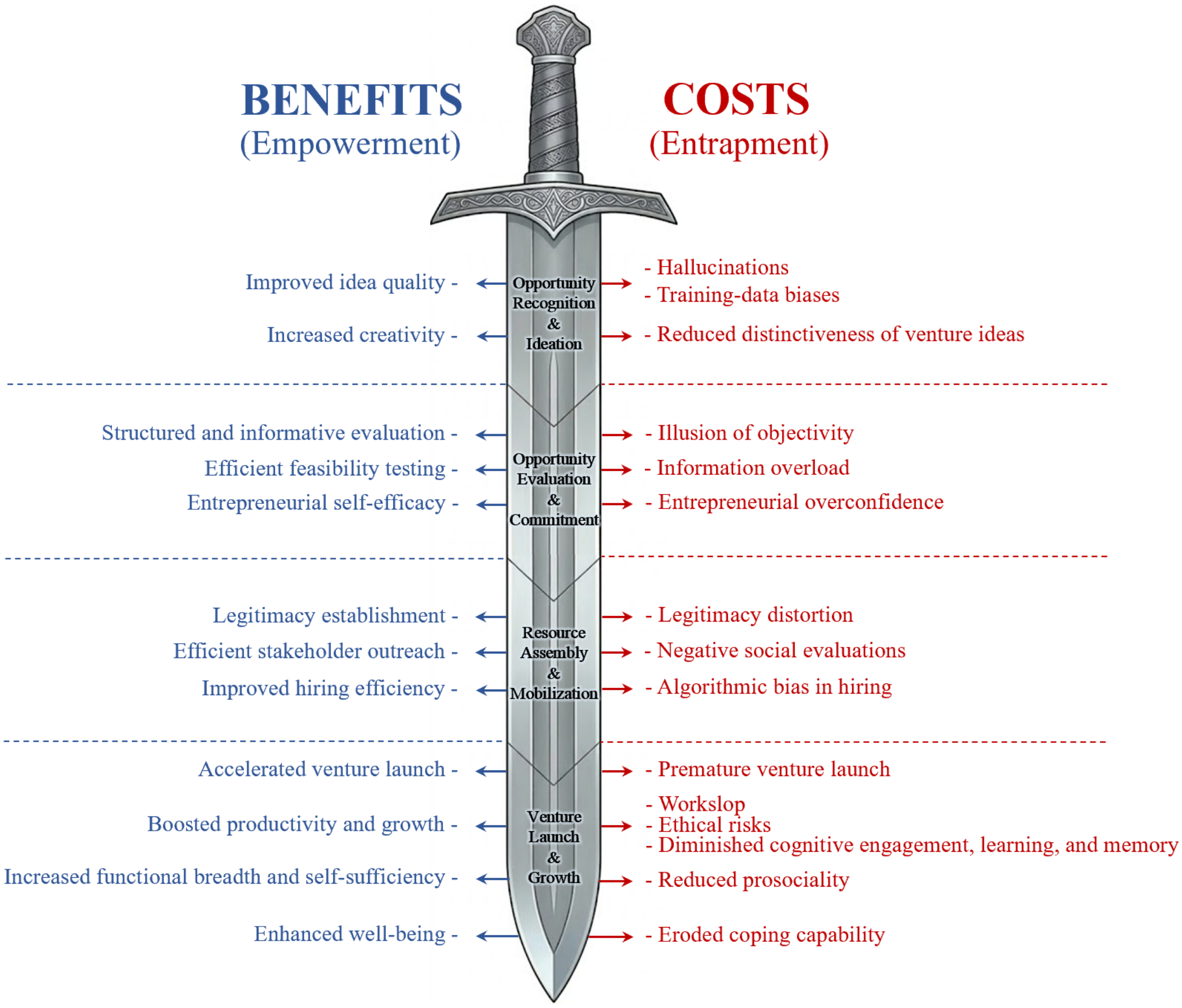

*Note.* The Empowerment–Entrapment Framework highlights the benefits and costs of GenAI that are most directly tied to the focal entrepreneurial task of each stage, while acknowledging that some effects recur across multiple stages of the entrepreneurial process.

**Table 1. Four Stages of the Entrepreneurial Process**

| Stage | Key entrepreneurial tasks |
|---|---|
| Opportunity recognition and ideation | Identify unmet needs and generate venture ideas |
| Opportunity evaluation and commitment | Assess whether to pursue, defer, or abandon the opportunity |
| Resource assembly and mobilization | Secure resources, legitimacy, and stakeholder support |
| Venture launch and growth | Bring the offering to market and build operating capabilities for growth |

*Note.* For analytical clarity, the framework presents the entrepreneurial process as a sequence of stages, consistent with stage-based process models (Baron, 2007b; Baron & Shane, 2007). In practice, however, the entrepreneurial process may unfold in a nonlinear, iterative manner, as emphasized in the effectuation literature and related perspectives (e.g., Bhave, 1994; Leatherbee & Katila, 2020; Sarasvathy, 2001).

**Table 2. Summary of How Generative AI’s Features Produce Benefits and Costs for Entrepreneurs**

| **GenAI Features** | **Corresponding Benefits** | Potential Explanation | **Corresponding Costs** | Potential Explanation |
|---|---|---|---|---|
| **The ability to complete tasks efficiently and conveniently** | Boosted productivity and growth<br>Enhanced well-being<br>Efficient stakeholder outreach<br>Efficient feasibility testing<br>Improved hiring efficiency<br>Accelerated venture launch | GenAI reduces the time and resource demands of entrepreneurial tasks (e.g., feasibility testing, networking emails, hiring, well-being support). | “Workslop”<br>Ethical risks<br>Diminished cognitive engagement, learning, and memory<br>Eroded coping capacity<br>Negative social evaluations | GenAI’s convenience entraps entrepreneurs in overreliance. |
| | Entrepreneurial self-efficacy | GenAI empowers entrepreneurs to feel more capable. | Entrepreneurial overconfidence<br>Premature venture launch<br>Reduced prosociality | GenAI’s empowerment fosters an inflated sense of readiness and self-sufficiency. |
| **Training data and design features of GenAI:**<br>• Broad and vast information<br>• Reliance on potentially biased real-world training data<br>• Probabilistic content generation<br>• Engineered to produce structured and detailed outputs | Improved idea quality<br>Increased creativity<br>Legitimacy establishment<br>Increased functional breadth and self-sufficiency | Entrepreneurs are empowered by GenAI’s broad and vast information. | Hallucinations | GenAI generates plausible outputs based on probabilistic patterns, without guaranteeing factual accuracy. |
| | | | Training-data biases<br>Algorithmic bias in hiring | GenAI reflects and reproduces biases embedded in its training data. |
| | | | Reduced distinctiveness of venture ideas | GenAI models trained on overlapping corpora may produce similar outputs. |
| | Structured and informative evaluation | GenAI is engineered to generate structured and detailed outputs. | Illusion of objectivity<br>Information overload<br>Legitimacy distortion | GenAI’s structured and detailed outputs may appear objective and overwhelming for users. |